\documentclass[letter,twocolumn]{jpsj2} 

\newcommand{\beq}{\begin{equation}}
\newcommand{\eeq}{\end{equation}}

\newcommand{\eqlabel}[1]{\label{eq:#1}}
\newcommand{\figlabel}[1]{\label{fig:#1}}
\newcommand{\Fig}[1]{Fig.\ \ref{fig:#1}}

\newcommand{\deffig}[4]{
\begin{figure}[tb]
  \begin{center}
   \includegraphics[width=#3 \textwidth]{#2}
  \end{center}
  \caption{ \figlabel{#1} #4}
\end{figure}
}

\begin{document}


\title{
  Tree Approximation for Spin Glass Models
}

\author{Naoki Kawashima}

\inst{
  Institute for Solid State Physics, 
  University of Tokyo, Kashiwa, Chiba 277-8581, Japan
}

\abst{
An approximate numerical approach to spin models is proposed,
in which the original lattice is transformed into a tree.
This method is applied to the Edwards-Anderson spin glass model in
two and three dimensions.
It captures the characteristics of each individual sample
and reproduces various qualitative features of sample averaged quantities
similar to those that have been observed in previous Monte Carlo simulations.
For example, from the Binder parameter for various 
system sizes as a function of the temperature we obtain
$T \sim 1.0$ with $\nu \sim 1.85$, 
in reasonable agreement with previous Monte Carlo simulations.
The present approximation yields the trivial structure
for the overlap distribution function.
}

\kword{
spin glass, 
real-space renormalization, 
Migdal-Kadanoff approximation,
Cayley tree
}

\maketitle

\section{Introduction}

The spin glass problem was brought forth 
by experiments on random magnets \cite{Experiments}
and the mathematical model was subsequently proposed \cite{EdwardsAnderson}.
Since then, the simplicity of the model and the fundamental nature of 
the problem have been encouraging the investment of a great deal of effort
on its study.
While the existence of the spin-glass phase transition was established
by Monte Carlo simulations \cite{Ogielski,BhattY,KawashimaY}
for the three-dimensional case,
the absense of a phase transition in two dimensions was more difficult 
to conclude and it was settled\cite{Houdayer,HoudayerH}
after the three-dimensional problem was solved.
Concerning the nature of the spin glass phase in three dimensions,
early theoretical works produced two paradigms that are
consistent with experimental findings:
the mean-field picture\cite{meanfield} and the droplet picture\cite{droplet}.
Therefore, investigations in the last two decades
have been focused on the question as to which paradigm is correct for 
spin glass models in two and three dimensions.
For solving this problem, the principal tool was numerical calculations.
It did not take too long, however, 
for researchers to realize the difficulty of 
the problem from the computational point of view.
Even a mathematically rigorous statement was made\cite{Barahona}:
the problem of finding the ground state of a given sample of the
Edwards-Anderson spin glass model is NP-hard if the lattice is
three- or higher-dimensional.

While this mathematical fact does not necessarily exclude the
possibility of finite-temperature calculation within a reasonable
amount of computational time,
no numerical method has proved to be polynomial-time bounded,
in particular, for temperatures below the critical point.
To this day, the largest size of the three-dimensional system 
of which the equilibrium properties can be studied within a realistic 
computational time does not exceed $L=30$,
which is not enough to settle the various important
issues beyond reasonable doubt.
The situation is rather discouraging because it is numerically
hopeless to push much further along the same line of effort
if the difficulty is a direct consequence of the NP-hardness 
of the ground state problem, as facts seem to point.

In the present article, therefore, we would like to call a renewed 
attention to another method for approaching the issue, i.e., 
an approximate numerical method.
One of the most well-known examples for methods of this kind is
the Migdal-Kadanoff (MK) real-space renormalization,\cite{MigdalKadanoff}
which yields a reasonably accurate result for the ferromagnetic Ising model
and was applied later to spin glasses.\cite{YoungS,SouthernY}
The procedure is often associated with the domain-wall renormalization
group\cite{McMillanI,McMillanII,BrayM} in which the domain-wall excitation 
free energy is identified with the effective coupling in
the renormalized scale.
It is, then, related to the droplet picture\cite{droplet} by associating
the domain-wall excitation to a droplet of the same scale.

\deffig{Contraction}{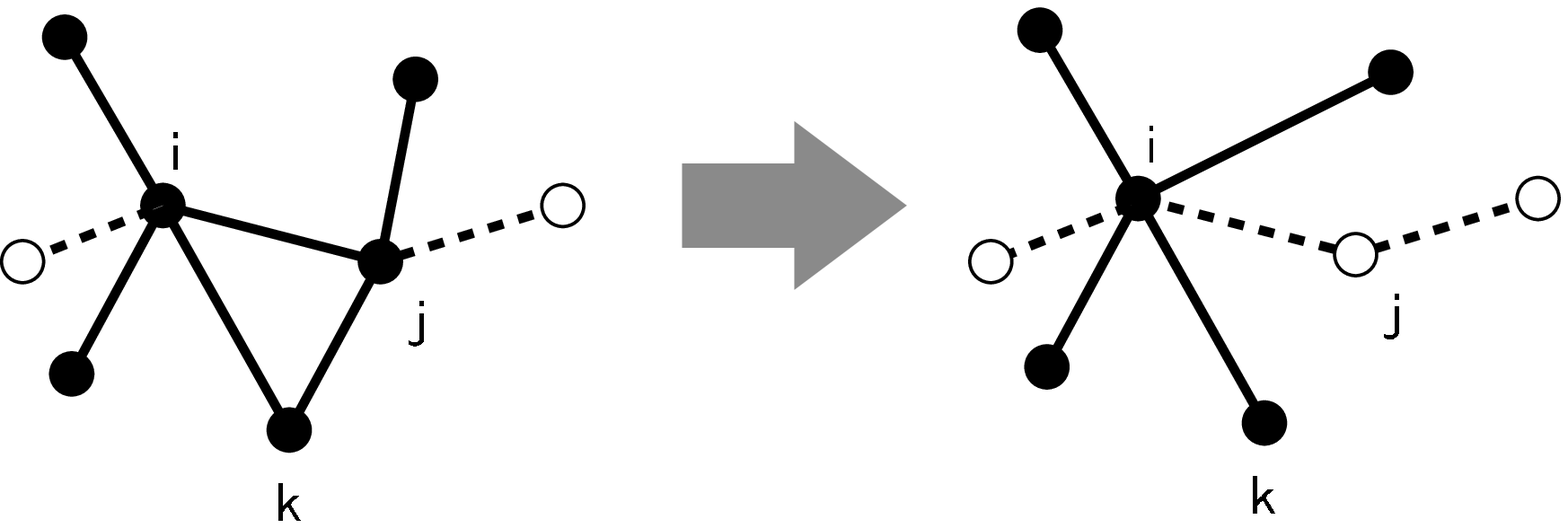}{0.45}{
Contraction process.
Circles are vertices and lines are edges.
Solid lines are those which are still in the list and
dashed lines are the ones that have been `contracted' and
eliminated from the list.
}

The procedure of the MK renormalization consists of two steps:
bundling bonds and tracing spins out.
The approximation is intuitively justified for the 
ferromagnetic system up to the correlation length
because in a region of a size smaller than the correlation length
most spins are parallel to each other and therefore 
the effect of $n$ parallel couplings in that region 
would be the same as a single stronger coupling 
that connects two neighboring regions.
Since this is not even qualitatively true for the spin glass systems
due to their high inhomogeneity,
the validity of the approximation is somewhat more
questionable there.
Nonetheless, the approximation seems to work well for spin glasses
\cite{YoungS,SouthernY}
judging from the fact that it produces results consistent with
other numerical findings, such as
the (likely) absence of phase transition in two dimensions,
\cite{McMillanI,McMillanII}
and its existence in three dimensions.
\cite{KawashimaY}

A disadvantage of the MK procedure, however, 
is that one cannot obtain results for a finite system
with a given bond configuration, say, in three dimensions.
This is due to the lack of direct correspondence between the
original spins in finite dimensions and the spins in the MK procedure.
This is one of the motivations that drive us to investigate
a new procedure while the main motivation is to improve the 
quality of the approximation taking into account the 
inhomogeneity of the system.

\section{Procedure}

We start from the graph of the same structure 
as the finite lattice on which the model is defined.
The vertices of the graph correspond to spins and edges to bonds.
We gradually deform the graph by contraction processes
discussed below until it becomes a tree with no loop.
Before starting the contraction processes, a vertex initially carries 
the volume of unity and an edge carries the weight equal to
the coupling constant of the corresponding bond.
The volumes of vertices and the weights of edges may be
changed during a contraction process.
All the edges are initially enlisted as the candidates for
contraction.

\deffig{ComparisonDQ1}{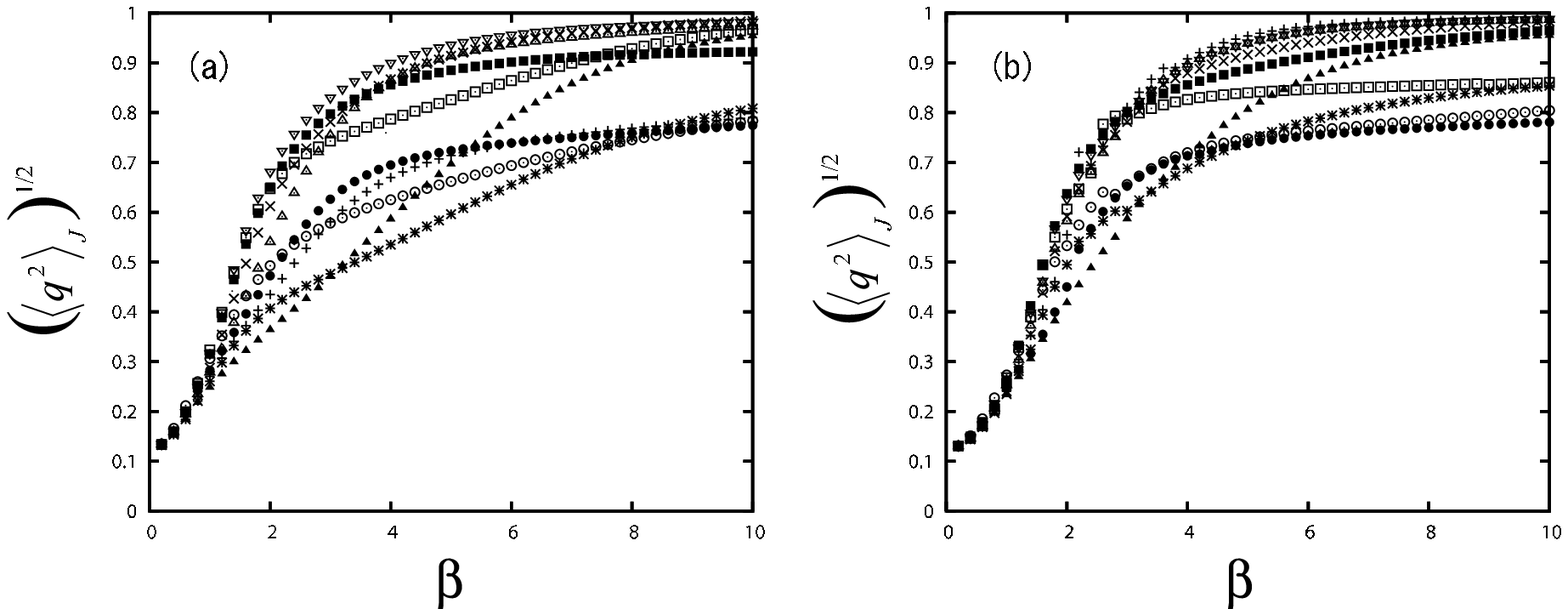}{0.48}{
The standard deviation of the overlap distribution function, $P_J(q)$,
for 10 randomly generated samples of an $8\times 8$ system.
The exact values are on the left panel, (a), and the 
result of the present approximation is on the right, (b).
The same bond samples are shown with the same symbols
in both the panels.
}

\deffig{ComparisonDQ2}{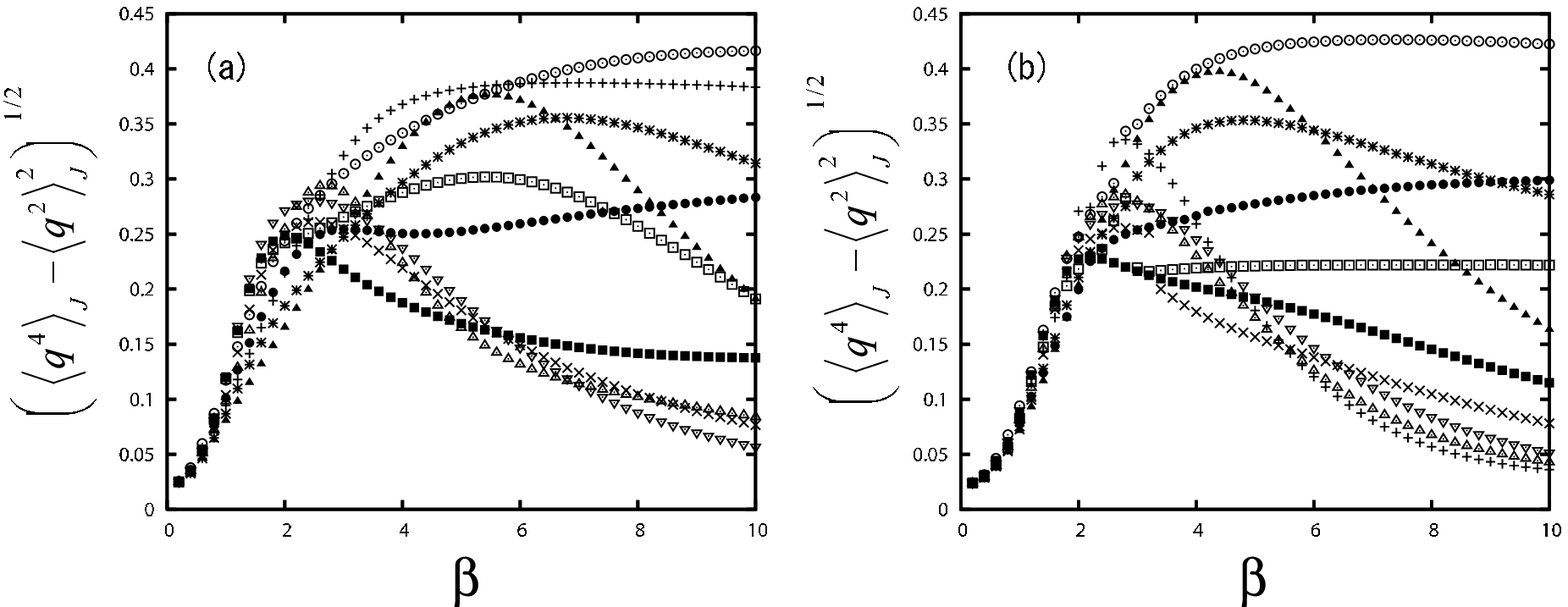}{0.48}{
The standard deviation of the squared overlaps.
The same samples as Fig.\,1.
}

The graph deformation is accomplished by 
applying the following contraction procedure repeatedly to 
the {\it heaviest} edge in the list until the list becomes empty.
Here, the heaviest edge is the one with the weight of 
the largest absolute value.
Let $e$ be the edge and 
$i$ and $j$ the two vertices connected by $e$,
where $i$ has a larger volume than $j$.
Then, we let $i$ enslave $j$.
This is done by detaching all the edges
connecting $j$ to other vertices, except $e=(ij)$,
and reconnect them to $i$. (\Fig{Contraction})
For a vertex $k$ that is 
connected to $j$ before the reconnection, the new weight
of the edge $(ik)$ after reconnection is
$
  J_{ik}^{\rm new} \equiv J_{ik}^{\rm old} + J'_{ik},
$
where $J'_{ik}$ is defined by
\begin{equation}
  \tanh (\beta J'_{ik}) \equiv 
  \tanh (\beta J_{ij}) \times \tanh (\beta J_{jk})
\end{equation}
with $\beta$ being the inverse temperature.
(If $i$ and $k$ are not directly connected before the
reconnection, $J_{ik}^{\rm old}$ is simply zero.)
The volume of the vertex $i$ is updated simply as
$
  V_i^{\rm new} \equiv V_i^{\rm old} + V_j^{\rm old}.
$
Then, the edge $e$ is removed from the list,
which is the end of the contraction procedure of the edge $e$.

By every contraction process, the number of edges is reduced
at least by one, and by definition of the procedure
the contracted part (consisting of dashed lines and
open circles in \Fig{Contraction}) does not have loops.
Therefore, when all the edges in the list have been contracted,
we are left with a tree.
Once we obtain the tree,
the spin-spin correlation function can be computed as 
the product of ``nearest-neighbor'' correlation functions
along the path leading from one spin to the other:
\begin{equation}
  \Gamma_{ij} \equiv \langle S_i S_j \rangle 
  =  \prod_{e \in \pi_{i\to j}} \tanh (\beta \tilde J_e),
  \eqlabel{Correlation}
\end{equation}
where $\pi_{i\to j}$ is the path on the tree 
from $i$ to $j$, and $\tilde J_e$ is the weight 
of the edge $e$ on the tree.
From correlation functions, we can obtain various
quantities of interest, such as the spin glass susceptibility
and the Binder parameter, as we present in what follows.
\cite{Note}

\deffig{Overlap2D}{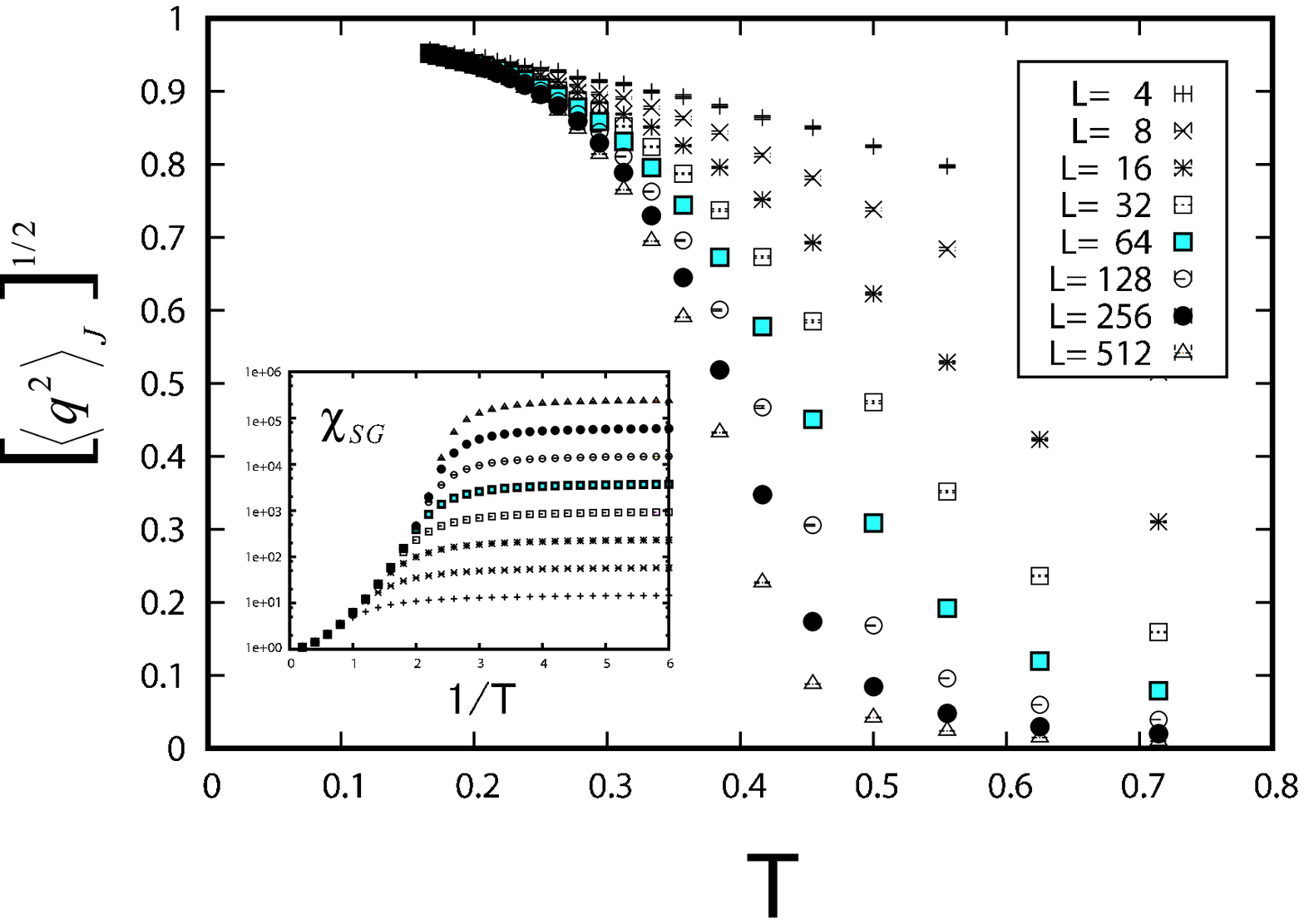}{0.42}{
The mean overlap $\sqrt{[\langle q^2 \rangle_J]}$ in two dimensions.
The inset is the spin glass susceptibility 
$\chi_{\rm sg} \equiv L^d [\langle q^2 \rangle_J]$
in the logarithmic scale.
($L=4$ and $8$ are not shown in the main panel.)
}

Here, a few remarks on the validity of the approximation may be due.
We should first note that the procedure yields exact results in one dimension.
In two or higher dimensions, we must treat the results of the 
approximation with greater caution.
Generally, we expect the present procedure to produce more accurate
results when the distribution function of the couplings is broader.
An approximation similar to the present one was proposed
for quantum systems \cite{Ma}.
It was shown to be {\it asymptotically exact}
for a few one-dimensional systems \cite{FisherRG} 
and applied also to two-dimensional systems \cite{Motrunich,Lin}.
This was possible because the width of the distribution diverges 
as we proceed in the renormalization transformation.
Since the same is not true in the present case,
the judgment of the validity or the utility of 
the present approximation must be based on what we 
actually obtain from it.
We may, however, say that the present scheme is expected to
be more accurate in lower dimensions when compared at the same
temperature, because in order for the present approximation 
to be accurate the largest bond must be dominating
among those which share the same spin,
the situation occurring less likely for a higher coordination number.

\section{Individual Samples}

In order to obtain a rough idea as to what degree the approximation 
is accurate for each individual bond sample, 
we first compute the variances of the overlap $q$ and
the squared overlap $q^2$, where $q$ is defined as
$
  q \equiv \frac1N \sum_{i=1}^N S_i^{(1)} S_i^{(2)}
$
with two replicas with the same bond configuration but
independent spins.
(The symbol $S_i^{(\alpha)}$ denote a spin at the site $i$
of the $\alpha$-th replica.)
Of crutial importance is the distribution of $q$,
\begin{equation}
  P_J(q) \equiv 
  \left\langle 
  \delta\left( q - \frac1N \sum_{i=1}^N S_i^{(1)} S_i^{(2)} \right) 
  \right\rangle_J,
\end{equation}
with $J$ specifying a particular bond configuration,
and its average over $J$, i.e., 
$
  P(q) \equiv [P_J(q)].
$
The square brackets $[\cdots]$ denote the average over
bond configurations hereafter.

The distribution $P_J(q)$ having a nontrivial structure is 
regarded as a defining character of the
mean-field picture whereas in the droplet picture
$P_J(q)$ is supposed to be trivial, i.e., 
a superposition of two delta functions at symmetric positions.
In both pictures, the spin glass phase is characterized by
the finite variance of $P_J(q)$.
Therefore, the phase transition from 
the high-temperature paramagnetic phase to the low-temperature
spin glass phase, can be detected in the behavior of
the variance of $P_J(q)$.
In \Fig{ComparisonDQ1}, the standard deviation is shown for
ten randomly chosen bond samples of two-dimensional
EA spin glass model.
The bond distribution is a Gaussian
with a mean value of zero and a variance of unity.

\deffig{Overlap3D}{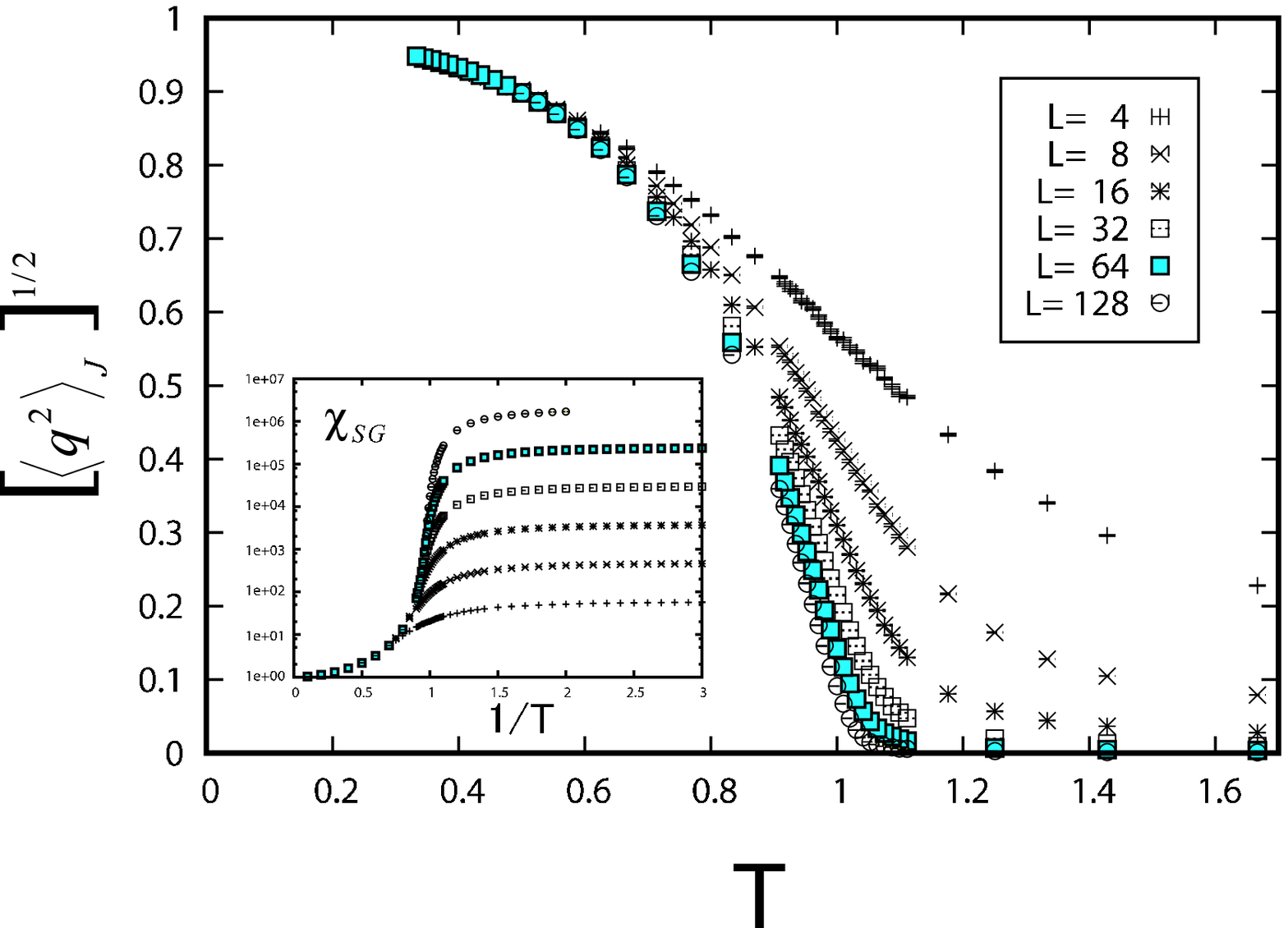}{0.42}{
The same as \Fig{Overlap2D} for the three-dimensional case.
}

The nature of the low-temperature phase may be better
reflected in the behavior of the variance of the
distribution of the squared overlap $q^2$, 
because the finite variance 
$\langle q^4 \rangle_J - \langle q^2 \rangle_J^2$
would be the most direct measure of the nontriviality of $P_J(q)$.
The estimates of the quantity are shown in \Fig{ComparisonDQ2}
for the same 10 bond samples as in \Fig{ComparisonDQ1}.

Looking at the two sets of figures, 
we notice that the agreement between the left panels (exact
calculation) and the right panels (approximation)
is markedly good for a crude approximation like the present one.
The locations of peaks and inflexion points in the right
panel (present approximation) roughly agree with those
in the left panel (exact) for most samples,
suggesting that the approximation captures 
the characteristic features in the temperature profile.

\section{Sample Averages}

\deffig{BinderParameter3D}{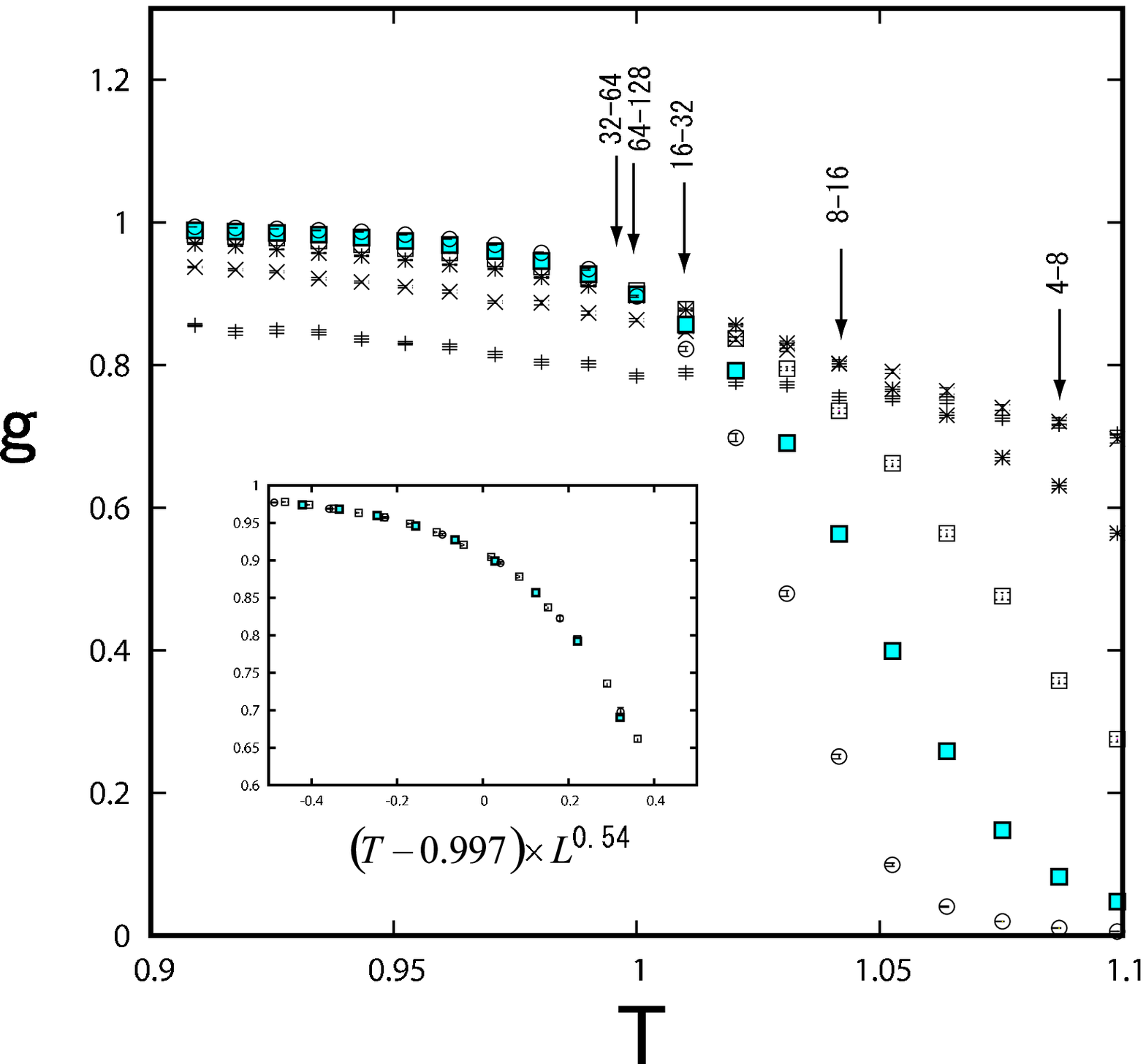}{0.42}{
  The Binder parameter for the three-dimensional model
  of system sizes of $L=4, 8, 16, 32, 64$ and $128$ (from top to
  bottom at high temperature).
  The arrows indicate the locations of the crossings between two
  successive system sizes.
  The inset shows the finite size scaling for the 
  largest three sizes, $L=$ 32, 64 and 128. (The other
  sizes are not shown in the inset.)
}

We now turn to the behaviors of averaged quantities.
In \Fig{Overlap2D}, we plot the standard deviation of $P_J(q)$
averaged over bond configurations for the two-dimensional system.
The inset is the spin glass susceptibility
$\chi_{\rm sg} \equiv N [\langle q^2 \rangle_J]$ against the
inverse temperature $\beta$.
The envelope function of $\chi_{\rm sg}$ in the region $\beta < 1$
shows roughly the same behavior as the previous computation\cite{HoudayerH},
which is free from any approximation.
Whereas the zero temperature phase transition
was concluded in the previous exact computation, 
it is difficult to judge from the present figure 
whether the transition temperature is finite or not.
The envelope function could be fit by a function for which the most 
singular term is $\exp(a \beta^2)$.
Equally plausible is a function that diverges around $\beta \sim 3.5$.
The previous exact computation shows that the susceptibility 
obeys the power law $\chi_{\rm sg} \propto T^{-2\nu}$,
where $\nu \sim 2.5$,
\cite{HoudayerH}
in the temperature range $1.5 < \beta < 2.5$.
In the same temperature range, the
present envelope function (in the double logarithmic scale) is steeper.
Therefore, we may not trust the critical exponent
estimated from the slope of the present envelope function.

For the three-dimensional system,
the spin glass susceptibility shows a larger change (\Fig{Overlap3D})
near $T \sim 1$, which is presumably the transition point.
In this case, we would have to assume an even more singular function
than $\exp(a \beta^2)$ in order not to allow a divergence 
at a finite temperature.
It seems much more natural to assume a finite temperature transition,
as is consistent with Monte Carlo simulations.\cite{Ogielski,BhattY,KawashimaY}
In \Fig{BinderParameter3D}, the Binder parameter,
$
  g \equiv [ ( 3 - 
  \langle q^4 \rangle_J / \langle q^2 \rangle_J^2 ) ] / 2,
$
is plotted against the temperature.
Curves of various system sizes cross each other and 
the crossing between larger systems takes place at lower temperature,
as is the case with the two-dimensional model (not shown).
However, the location seems to converge to some finite value in the
thermodynamic limit, in contrast the two-dimensional case.
In particular, the value of $g$ at the crossings increases in $g$
up to the $L \sim 32$, while we do not observe a statistically 
significant increase above this scale.
The crossing point in the thermodynamic limit is located
around $T \sim 1.0$ and $g \sim 0.9$.
We can also produce a finite-size-scaling plot
that looks reasonable, as shown in the inset of \Fig{BinderParameter3D}.
The plot yields the estimates of the critical exponent; $\nu \sim 1.85$.
The agreement of these values with the results of Monte Carlo simulation,
$T_c = 0.98(5)$ and $\nu = 2.00 (15)$,
\cite{MarinariPR} are markedly good considering that the
crudeness of the present approximation.
This is also an improvement on the MK result
\cite{SouthernY},
$T_{\rm c} \sim 0.39$ and $\nu \sim 2.8$.

\deffig{Variance3D}{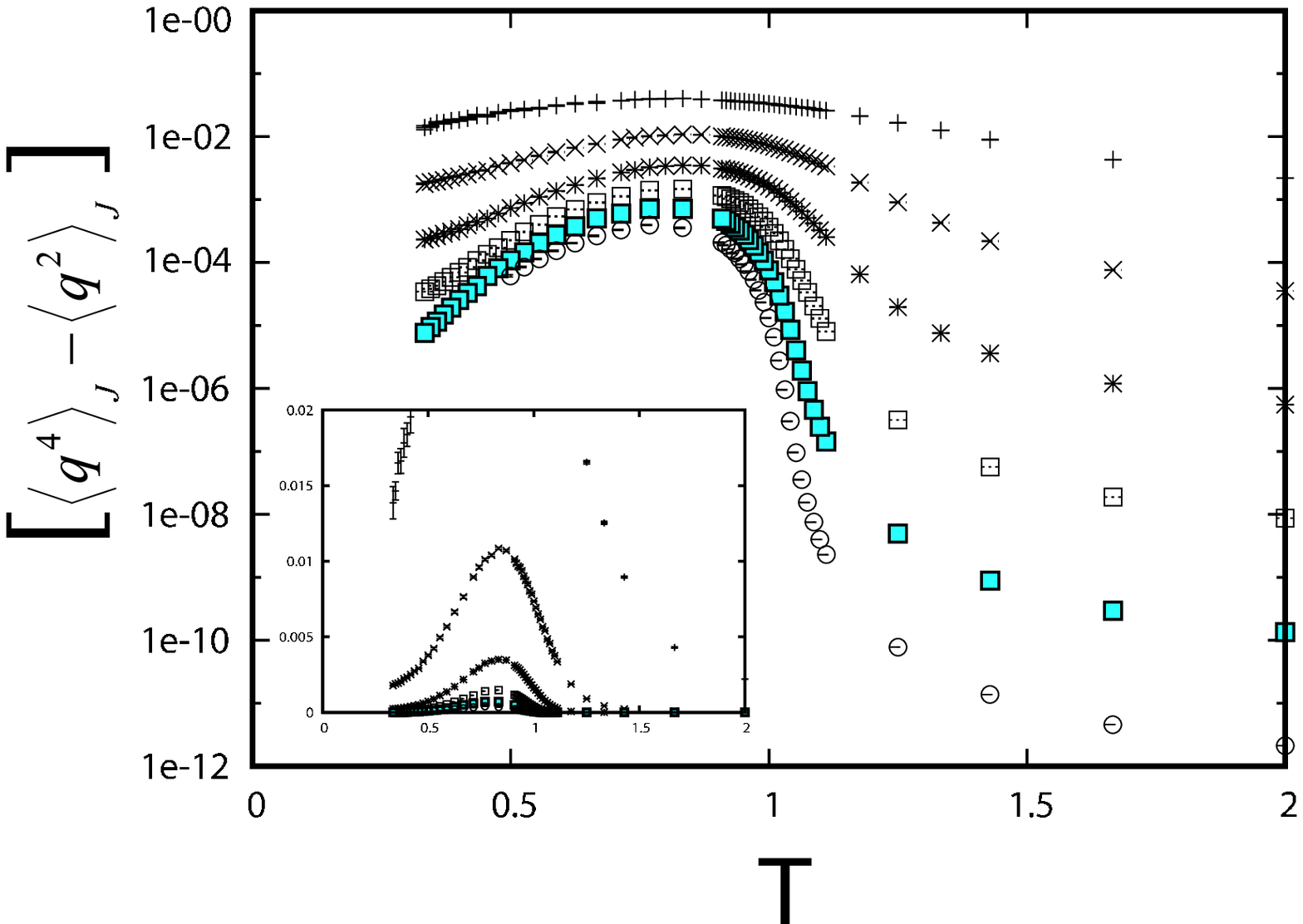}{0.42}{
  The variance of the distribution of $q^2$ 
  for the three-dimensional model of $L=4, 8, 16, 32, 64$ and $128$
  (from top to bottom) in the logarithmic scale.
  The same quantity is plotted in the linear scale in the inset.
}

In \Fig{Variance3D}, the variance of $q^2$, averaged over bond 
configurations, is shown in a logarithmic scale.
(The same is presented in a linear scale in the inset.)
The variance is a decreasing function of the system size
at any temperature.
The decrease is rapid both at the high and low temperature ranges
and is somewhat more moderate in the middle.
However, even at the temperature where the system size 
dependence is the weakest, the variance is extremely small
and seems to converge to zero in the thermodynamic limit,
consistent with the droplet scenario.


\section{Concluding Remarks}

We have proposed a simple procedure that is motivated by the
real-space renormalization-group methods of MK
and analogous to that of Dasgupta, Ma and Hu\cite{Ma}.
It produces a self-consistent framework for understanding 
spin glass systems and makes it possible to compute 
various quantities, including $P(q)$, for a large given sample
in finite dimensions, which is otherwise impossible to compute.
The results for three-dimensional systems
show an unexpectedly good agreement with the other numerical
methods that do not involve any approximation, while
the low-temperature behaviour in two dimensions shows the
limitation of the present approximation.
As for the triviality/nontriviality of $P(q)$ in the 
spin glass phase, the approximation supports the trivial structure.
This is natural, although not obvious a priori, 
considering that the present method is similar to the MK procedure.
Therefore, the present finding should not be regarded as 
new evidence for the validity of the droplet picture.
Rather, the present method should be used as a guide for 
other numerical methods with no approximation, such as
Monte Carlo simulation, as one can easily draw a wrong
conclusion from those methods because of severe limitation
on the system size and relatively large corrections to scaling.
The present result suggests that large systems (say, $L > 32$) may
be required in the three-dimensional case in order to get rid of
the correction to scaling and obtain a reliable estimate of 
critical exponents.
Another advantage of the present method,
that it may enable us to compute dynamical quantities 
much more easily than using other methods, 
will be exploited in a future study.

A part of the numerical calculation in this work has been carried out 
using the facilities of the Supercomputer Center, 
Institute for Solid State Physics, University of Tokyo.


\begin{thebibliography}{99}
\bibitem{Note} 
Four-point functions, which we need for the Binder parameter,
can be reduced into products of two-point functions,
analogous to Wick's theorem for the Gaussian integration.
\end{thebibliography}
\end{document}